\documentclass[a4paper,10pt]{article}
\usepackage[utf8x]{inputenc}
\usepackage{amsmath, amssymb, amsfonts}
\usepackage{amsthm}
\newtheorem{lemma}{Lemma}
\newtheorem{hilfsatz}{Hilfsatz}
\newtheorem*{main_th}{Main Theorem}

\title{Tilings With Very Elastic Dimers}
\author{Paul Federbush \\
Department of Mathematics \\
University of Michigan \\
Ann Arbor, MI 48109-1043 \\
(pfed@umich.edu)}

\begin{document}

\maketitle

\begin{abstract}
We consider tiles (dimers) each of which covers two vertices of a rectangular lattice.  There is a normalized translation invariant weighting on the shape of the tiles.  We study the pressure, $p$, or entropy, (one over the volume times the logarithm of the partition function).  We let $p_0$ (easy to compute) be the pressure in the limit of absolute smoothness (the weighting function is constant).  We prove that as the smoothness of the weighting function, suitably defined, increases, $p$ converges to $p_0$, uniformly in the volume.  It is the uniformity statement that makes the result non-trivial.  In an earlier paper the author proved this, but with an additional requirement of a certain fall-off on the weighting function.  Herein fall-off is not demanded, but there is the technical requirement that each dimer connect a black vertex with a white vertex, vertices colored as on a checker board.  This seems like a very basic result in the theory of pressure (entropy) of tilings. 
\end{abstract}

\section*{Introduction}
In a previous work, \cite{Fed1}, we studied tilings by tiles of arbitrary size.  That is, each tile therein covered $n$ vertices.  Here we specialize to $n=2$.  We here require that each of our dimers (size equals two tiles) cover one black vertex and one white vertex.  There was no such requirement in \cite{Fed1}.  The proof in \cite{Fed1} required both smoothness and fall-off conditions in the limit that drove $p$ to $p_0$.  Here we need control only the smoothness.  We believe strongly only smoothness conditions should be needed to yield the convergence in the generality of the problem treated in \cite{Fed1}!  This seems an interesting challenge, to find such a proof.  

In proving the theorem of this paper, we obtain the existence of the limit of $p$ by employing lower and upper bounds on $p$.  The upper bound we use comes directly from \cite{Fed1}, but we repeat the proof of this bound here, so that this paper may be read independently of \cite{Fed1}.  The condition that each dimer cover one black vertex and one white vertex is imposed so that we may identify the partition function with the permanent of a certain matrix.  We will explain this below for those unfamiliar with this standard formalism.  In this setting a very basic lower bound on the permanent, \cite{Friedland}, suffices for our purpose.  So this paper could have been written as a one page paper quoting results from \cite{Fed1} and \cite{Friedland}!

The idea of studying partition functions of tilings, with a weighting function on the tiles, in the limit the weighting function becomes smoother and smoother, has had a very fruitful outcome.  Formally, the weighting functions yielding traditional dimers become smoother and smoother as the dimension is increased.  In this case the weighting function is one on a dimer covering two nearest neighbor vertices, and zero on all other shaped dimers.  (It is not obvious that such a weighting formally gets smoother with dimension.)  Following this line, there has been developed a presumed asymptotic expansion for $\lambda_d$ of the dimer problem in $d$-dimensions, \cite{Fed2}.

\section*{Formalism}
We work with a $d$-dimensional cubic lattice torus, $\Lambda$, of edge size $L$.  We view this as either
\begin{align}
\left( \mathbb{Z} / L \mathbb{Z} \right)^d
\end{align} 
or as the subset of $\mathbb{Z}^d$ given by
\begin{align}
\label{subset}
\left\{0, \dots , L-1 \right\}^d.
\end{align}
We let the number of vertices of $\Lambda$ be $2N$,
\begin{align}
2N = L^d.
\end{align}
We \textit{color} the lattice, the points of $\Lambda$ the sum of whose coordinates are even are black, the sum of whose coordinates are odd are white.  We require $L$ to be even, so this makes sense.  

The \textit{weighting function}, $f \left( x,y \right)$, is a symmetric function on $\Lambda \times \Lambda$.  Viewing $\Lambda$ as a subset of $\mathbb{Z}^d$ via (\ref{subset}), we extend $f$ to a periodic function on $\mathbb{Z}^d \times \mathbb{Z}^d$ requiring
\begin{align}
f \left( x+c , y+d \right) = f \left( x,y \right) 
\end{align}
where $c$ and $d$ are points in $\mathbb{Z}^d$ all of whose coordinates are divisible by $L$.  The weighting function is required to be \textit{positive},
\begin{align}
f \left( x,y \right) > 0
\end{align}
and \textit{translation invariant},
\begin{align}
\label{translation}
f \left( x+a , y+a \right) = f \left( x,y \right)
\end{align}
for each point $a$ in $\mathbb{Z}^d$.  (In (\ref{translation}) and other similar equations it is important to remember we've extended $f$ to all values of $\mathbb{Z}^d \times \mathbb{Z}^d$.)  In addition $f$ is \textit{normalized} by
\begin{align} 
\label{normalize_f}
\sum_{y \perp x} f \left( x,y \right) = 1.
\end{align}
In the summation $x$ is fixed and $y$ is summed over all the $y \in \Lambda$ of color opposite to that of $x$, the $\perp$ sign indicating opposite color.

We measure the \textit{smoothness} of $f$ by defining $\mathrm{sm} \left( f \right)$ to be chosen as the smallest value satisfying
\begin{align}
\label{sm_boundf}
\left| f \left( x,y \right) - f \left( x, y+u \right) \right| \le \mathrm{sm} \left( f \right) \cdot f \left( x,y \right)
\end{align}
for all $x, y,$ and $u$.  $u$ is a unit vector, of which there are $2d$.  

We now proceed to define the \textit{partition function}, $Z \left( f \right)$, a function of the weighting function, $f$.  We first define a \textit{tiling} of $\Lambda$ to be a set of two-element subsets of $\Lambda$
\begin{align}
T_i = \left\{ s_1^i, \dots, s_N^i \right\}
\end{align}
where the $s^i_j$ are disjoint (and so cover $\Lambda$), and each $s^i_j$ contains one black and one white vertex.  We write
\begin{align}
s_j^i = \left\{ x_j^i, y_j^i \right\}.
\end{align}
The partition function is then given by
\begin{align}
Z \left( f \right) = \sum_{T_i} \prod_{s_j^i \in T_i} f \left( x_j^i, y_j^i \right)
\end{align}
where the sum is over all distinct tilings.  (It is easy to see there are exactly $N!$ tilings.)  We define the pressure (entropy), $p$, associated to a partition function by
\begin{align}
e^{2Np} = Z.
\end{align}
If $f$ is infinitely (perfectly) smooth, and so constant (by (\ref{sm_boundf})), then by (\ref{normalize_f}) one has $f = f_0$,
\begin{align}
f_0 = \frac{1}{N}.
\end{align}
Then $Z \left( f_0 \right)$ is easily seen to be
\begin{align}
Z \left( f_0 \right) = \left( \frac{1}{N} \right)^N N!
\end{align}
and with
\begin{align}
e^{2Np_0} = Z \left( f_0 \right)
\end{align}
one has
\begin{align}
\label{p0}
p_0 = \frac{1}{2N} \left[ \ln N! - N \ln N \right].
\end{align}
And we take the infinite volume limit to get
\begin{align}
\label{p0bar}
\bar{p}_0 = \lim_{N \to \infty} p_0 = - \frac{1}{2}.
\end{align}

\section*{Results}
\begin{lemma}
\label{zbound_up}
For all $f$ one has
\begin{align}
Z \left( f \right) \le 1.
\end{align}
\end{lemma}
This is a special case of Lemma 1 of \cite{Fed1}, with in this special case, a better bound.  As we said before, we will present complete proofs herein, not depending on results from \cite{Fed1}.

\begin{lemma}
\label{zbound_lo}
For all $f$ one has
\begin{align}
Z \left( f \right) \ge \left( \frac{1}{e} \right)^N.
\end{align}
\end{lemma}
This is the basic result about permanents, to be explained later.

\begin{lemma} \textbf{The Root Estimate} \\
\label{root_est}
Let $a_{ij} \ge 0$ and $\left| \delta_{ij} \right| \le 1$, set
\begin{align}
\bar{\delta} = \max \left( \left\{ \delta_{ij} \right\} \right) \\
\hat{\delta} = \min \left( \left\{ \delta_{ij} \right\} \right)
\end{align}
and let $A$ be given by
\begin{align}
A = \left( \sum_{i} \prod_{j=1}^N a_{ij} \right)^{\frac{1}{N}}.
\end{align}
Then
\begin{align}\
\label{deltas_bound}
\left( 1 + \hat{\delta} \right) A \le \left( \sum_i \prod_{j=1}^N a_{ij} \left( 1 + \delta_{ij} \right) \right)^{\frac{1}{N}} \le \left( 1 + \bar{\delta} \right) A.
\end{align}
\end{lemma}
This is stated following Lemma 2 of \cite{Fed1}.

\begin{lemma}
\label{partition_bound}
Let $f_1$ and $f_2$ be two weighting functions, and assume, with $\epsilon < 1$,
\begin{align}
\left| f_1 - f_2 \right| \le \epsilon f_1
\end{align}
where this is a pointwise bound.  Then
\begin{align}
\left( 1 - \epsilon \right) \left( Z \left( f_1 \right) \right) ^{\frac{1}{N}} \le \left( Z \left( f_2 \right) \right)^{\frac{1}{N}} \le \left( 1 + \epsilon \right) \left( Z \left( f_1 \right) \right)^{\frac{1}{N}}.
\end{align}
\end{lemma}
This is a form of Lemma 2 of \cite{Fed1}.

\begin{main_th}
For each $\epsilon > 0$ there is a $\delta = \delta \left( \epsilon \right)$ such that
\begin{align}
\left| p \left( f \right) - p_0 \right| < \epsilon
\end{align}
if
\begin{align}
\mathrm{sm} \left( f \right) < \delta.
\end{align}
\end{main_th}  
$p \left( f \right)$ and  $p_0$ are also functions of $N$, but $\delta$ may be picked independent of $N$.

The $\delta$ we will find for given $\epsilon$ is determined in part by some nonconstructive processes.  The line of proof limits its size in a number of demanding steps.  One believes the $\delta$ that work should be given by simpler conditions.  Finding \textit{realistic} choices of $\delta$ remains a challenge.  It is certainly related to finding a better proof, likely one not requiring the separation into black and white vertices.

This paper contains two main ideas.  The first is the root estimate and the definition of smoothness, eq. (\ref{sm_boundf}), chosen to dovetail with the root estimate for applications.  The second idea is the definition and application of $\bar f$ below.  Beyond these two ideas the rest is technical complication and some hard work.  In fact, then, both main ideas are present in \cite{Fed1} in a more complex form, since there the tiles may cover more than two vertices.

\section*{Proofs, I}

\textbf{Proof of Lemma \ref{zbound_up}.}  We observe
\begin{align}
Z \left( f \right) &= \sum_{T_i} \prod_{s_j^i \in T_i} f \left( x_j^i , y_j^i \right) \notag \\
&\le \prod_{{x \atop x \ \text{black}}} \left( \sum_{y \perp x} f \left( x,y \right) \right) = 1.
\end{align}
\qed

\textbf{Proof of Lemma \ref{zbound_lo}.}
We first relate $Z$ to a \textit{permanent}.  We consider a matrix $M$, $N \times N$, whose rows are labelled by the black vertices of $\Lambda$ and whose columns are labelled by the white vertices of $\Lambda$.  We set $M_{ij} = f \left( i,j \right)$.  Then all the entries of $M$ are non-negative, and the sum of the entries in each row, and likewise in each column, is 1.  Such a matrix is called \textit{doubly stochastic}.  The \textit{permanent} of $M$ is the sum of the same terms as defining the \textit{determinant} of $M$, but with all the minus signs in the definition of the determinant changed to plus signs.  And one easily sees
\begin{align}
Z \left( f \right) = \mathrm{Permanent} \left( M \right).
\end{align}
The fundamental theorem of \cite{Friedland}, that 
\begin{align}
\mathrm{Permanent} \left( M \right) \ge \left( \frac{1}{e} \right)^N
\end{align}
yields the lemma.  Reference \cite{Minc} is a standard reference on permanents.

\textbf{Proof of Lemma \ref{root_est}.}
One has immediately from 
\begin{align}
\left( 1 + \hat{\delta} \right) a_{ij} \le a_{ij} \left( 1 + \delta_{ij} \right) \le \left( 1 + \bar{\delta} \right) a_{ij}
\end{align}
and the fact that all these terms are positive the statement (\ref{deltas_bound}).
\qed

\textbf{Proof of Lemma \ref{partition_bound}.}
We write
\begin{align}
f_2 = f_1 + \left( f_2 - f_1 \right)
\end{align}
and let
\begin{align}
a_{ij} &= f_1 \left( i,j \right) \\
\delta_{ij} &= \left( f_2 \left( i,j \right) - f_1 \left( i,j \right) \right) / f_1 \left( i,j \right) \notag
\end{align}
then setting
\begin{align}
\bar{\delta} &\le \epsilon \\
\hat{\delta} &\ge - \epsilon
\end{align}
Lemma \ref{partition_bound} then follows from Lemma \ref{root_est}, in (\ref{deltas_bound})  $i$ is a black vertex and the sum over $j$ is over $j \perp i$.

\section*{Proofs, II, the Main Theorem}
We turn to the nitty-gritty, proving the Main Theorem.  We are given an $\epsilon > 0$, and we find the $\delta = \delta \left( \epsilon \right)$ that works in a number of steps.
\\ \\
\underline{Step 1.}  Find $N_1$ such that if 
\begin{align}
\left| \Lambda \right| = 2N > 2N_1
\end{align}
then 
\begin{align}
\label{ppbar-diff}
\left| p_0 \left( \Lambda \right) - \bar{p}_0 \right| < \frac{\epsilon}{2}.
\end{align}
Refer to (\ref{p0}) for $p_0$.  This is trivially possible by (\ref{p0bar}).
\\ \\
\underline{Step 2.}  Find $\delta_1$ such that if 
\begin{align}
\label{delta1}
\left| \left( Z \left( f \right) \right)^{\frac{1}{N}} - \frac{1}{e} \right| < \delta_1
\end{align}
then
\begin{align}
\label{eps/2}
\left| p \left( f \right) - \bar{p}_0 \right| < \frac{\epsilon}{2}.
\end{align}
Equation (\ref{delta1}) is equivalent from definitions to
\begin{align}
\left| e^{2p \left( f \right)} - e^{2\bar{p}_0} \right| < \delta_1.
\end{align}
By Lemmas \ref{zbound_up} and \ref{zbound_lo} all $p$'s satisfy
\begin{align}
-\frac{1}{2} \le p \le 0.
\end{align}
It is easy to show, using the mean value theorem, that the following choice works
\begin{align}
\delta_1 = \frac{\epsilon}{e}.
\end{align}

Since by Lemma \ref{zbound_lo}, $\left( Z \left( f \right) \right)^{1/N} \ge \frac{1}{e}$, we may replace the requirement (\ref{delta1}) by
\begin{align}
\left( Z \left( f \right) \right)^{\frac{1}{N}} < \frac{1}{e} + \delta_1.
\end{align}
\\ \\
\underline{Step 3.}  We choose $\bar{l}$ even and $\bar{n}$ integers with 
\begin{align}
2 \bar{n} = \bar{l}^d
\end{align}
such that 
\begin{align}
\label{nfacfrac-bound}
\left( \frac{\bar{n}!}{\bar{n}^{\bar{n}}} \right)^{1 / \bar{n}} < \frac{1}{e} + \frac{\delta_1}{4}.
\end{align}
We then take the realization of $\Lambda$ as given in (\ref{subset}) and divide it into cubes of edge size $\bar{l}$ and pieces of such cubes that are cut off at the boundary of (this realization of) $\Lambda$.  Let $\left\{ C_{\alpha} \right\}$ be this set of cubes and pieces of cubes.  We define $\bar{f} \left( x,y \right)$, for $x$ a black vertex and $y$ a white vertex, by
\begin{align}
\label{fbar}
\bar{f} \left( x,y \right) = \frac{1}{\# \left\{ y_i \mid y_i \sim y \right\} } \sum_{y_j \sim y} f \left( x, y_j \right). 
\end{align}
Here $y' \sim y$ if they are contained in the same $C_{\alpha}$ in $\left\{ C_{\alpha} \right\}$.  The right side of (\ref{fbar}) is the sum over values of $f \left( x,y_j \right)$ for those $y_j$ in the same $C_{\alpha}$ divided by the number of such $y_j$.  $\bar{f}$ is the \textit{average} of $f$ over a little cube (or part thereof), in the white vertex variable.
\\ \\
\underline{Step 4.}  We now present a special case of Lemma 3 of \cite{Fed1}.

\begin{hilfsatz}
\label{hilf_f_fbar}
Assume
\begin{align}
\alpha = \bar{l} d \cdot \mathrm{sm} \left( f \right) < 1
\end{align}
then
\begin{align}
\left| \bar{f} - f \right| \le \left( \frac{\alpha}{1 - \alpha} \right) f.
\end{align}
\end{hilfsatz}
We note that using this hilfsatz we can find a $\delta_3$ such that 
\begin{align}
\label{zzbar-diff}
\left| \left( Z \left( f \right) \right)^{\frac{1}{N}} - \left( Z \left( \bar{f} \right) \right)^{\frac{1}{N}} \right| < \frac{\delta_1}{2}
\end{align}
if
\begin{align}
\mathrm{sm} \left( f \right) < \delta_3.
\end{align}

\textbf{Proof of Hilfsatz \ref{hilf_f_fbar}.}  Let $M$ and $m$ be the maximum and minimum values of $f$ as the second variable wanders over a $C_{\alpha}$ (see (\ref{fbar})).  Then one has 
\begin{align}
\left| M - m \right| &\le \Sigma \Delta f \le \bar{l}d \cdot \mathrm{sm} \left( f \right) M \\
M - m &\le \alpha M \\
M - m &\le \frac{\alpha}{1 - \alpha} m \\
\left| f - \bar{f} \right| &\le \frac{\alpha}{1 - \alpha} f.
\end{align}
The implications above each follow upon some reflection and minor computation.
\\ \\
\underline{Step 5.}  Let $n_{\alpha}$ be the number of white vertices of $C_{\alpha}$.

\begin{hilfsatz}
\label{hilf_zbound}
\begin{align}
\label{z-fbar-bound}
Z \left( \bar{f} \right) \le \prod_{\alpha} \left( \frac{n_{\alpha}!}{n_{\alpha}^{n_{\alpha}}} \right).
\end{align}
Here the product is over the indices $\alpha$ that label the $C_{\alpha}$.
\end{hilfsatz}

\textbf{Proof of Hilfsatz \ref{hilf_zbound}.} This proof is sort of amusing, easy to see in your head once the idea is grasped.  We write
\begin{align}
Z \left( \bar{f} \right) \le \left( \prod_{i \ \mathrm{black}} \left( \sum_{j \ \mathrm{white}} \bar{f} \left( i,j \right) \right) \right) \cdot \prod_{\alpha} \left( \frac{n_{\alpha}!}{n_{\alpha}^{n_{\alpha}}} \right).
\end{align} 
The right side clearly is the right side of (\ref{z-fbar-bound}).  And if we restrict the product of sums, keeping only terms where the number of $j$'s landing in $C_{\alpha}$ is $n_{\alpha}$; then the restricted sum is the left side of (\ref{z-fbar-bound}).  The second product puts in each $j$ in $C_{\alpha}$ in a separate white vertex of $C_{\alpha}$.
\\ \\
\underline{Step 6.}  Let's look at the paradigm that shapes the proof.  Assume $\left| \Lambda \right| > \left| \Lambda_1 \right| =2N_1$ and that $\bar{l}$ divides $L$.  Then 
\\ \\
(a)  $n_{\alpha} = \bar{n}$ for all $\alpha$ and from (\ref{z-fbar-bound}) and (\ref{nfacfrac-bound})
\begin{align}
\left( Z \left( \bar{f} \right) \right)^{\frac{1}{N}} < \frac{1}{e} + \frac{\delta_1}{4}.
\end{align}
\\ \\
(b)  From (\ref{zzbar-diff})
\begin{align}
\left| \left( Z \left( \bar{f} \right) \right)^{\frac{1}{N}} - \left( Z \left( f \right) \right)^{\frac{1}{N}} \right| < \frac{1}{2} \delta_1, \ \text{if} \ \mathrm{sm} \left( f \right) < \delta_3
\end{align}
\\ \\
(c) so
\begin{align}
\left( Z \left( \bar{f} \right) \right)^{\frac{1}{N}} < \frac{1}{e} + \delta_1
\end{align}
\\ \\
(d)  and therefore by (\ref{delta1}) and (\ref{eps/2})
\begin{align}
\left| p \left( f \right) - \bar{p}_0 \right| < \frac{\epsilon}{2}
\end{align}
\\ \\
(e)  and from (\ref{ppbar-diff}) that
\begin{align}
\left| p \left( f \right) - p_0 \left( \Lambda \right) \right| < \epsilon.
\end{align}
The restrictions then are $\left| \Lambda \right| > \left| \Lambda_1 \right|$, $\bar{l}$ divides $L$, and
\begin{align} 
\mathrm{sm} \left( f \right) \le \delta_3 = \delta.
\end{align}
One has been given $\epsilon$, determining $\delta_1$ in terms of $\epsilon$, then determining $\bar{n}$ in terms of $\delta_1$, and then $\delta_3$ in terms of $\bar{n}$ and $\delta_1$ (in a rather complicated way).

All the cleverness is in this paradigm case.

It is just a little work to extend to the situation when $\bar{l}$ does not divide $L$.  The treatment of $\Lambda$ for which $\left| \Lambda \right| \le \left| \Lambda_1 \right|$ is easy.
\\ \\
\underline{Step 7.  Picking up the pieces, Checkmate.}  We treat the easier problem first.  Suppose there is a $\delta$ and $\Lambda_{\epsilon}$ such that if $\left| \Lambda \right| > \left| \Lambda_{\epsilon} \right|$ then
\begin{align} 
\left| p \left( f \right) - p_0 \left( \Lambda \right) \right| < \epsilon \ \text{if} \ \mathrm{sm} \left( f \right) < \delta.
\end{align}
It is trivial that for a given $\left| \Lambda \right|$, for each $\tilde{\epsilon}$ there is a $\tilde{\delta}$ such that 
\begin{align}
\left| p \left( f \right) - p_0 \left( \Lambda \right) \right| < \tilde{\epsilon} \ \text{if} \ \mathrm{sm} \left( f \right) < \tilde{\delta}.
\end{align}
So there is a $\bar{\delta}$ such that if $\left| \Lambda \right| \le \left| \Lambda_{\epsilon} \right|$
\begin{align}
\left| p \left( f \right) - p_0 \left( \Lambda \right) \right| < \epsilon \ \text{if} \ \mathrm{sm} \left( f \right) < \bar{\delta}.
\end{align}
Then $\hat{\delta} = \min \left( \delta , \bar{\delta} \right)$ will work for all $\Lambda$.

The last remaining problem is to deal with the $n_{\alpha} < \bar{n}$.  We let $N_1$ be the number of white vertices in cubes $C_{\alpha}$ with $n_{\alpha} < \bar{n}$, and $N_2$ be the number of white vertices in cubes $C_{\alpha}$ with $n_{\alpha} = \bar{n}$.  Then from (\ref{z-fbar-bound}) we see
\begin{align}
\label{z-fbar-bound2}
\left( Z \left( \bar{f} \right) \right)^{\frac{1}{N}} \le \left( \frac{1}{e} + \frac{\delta_1}{4} \right)^{\frac{N_2}{N}} \cdot \left( 1 \right)^{\frac{N_1}{N}}.
\end{align}
Note
\begin{align}
N_1 + N_2 = N = \frac{L^d}{2}
\end{align}
and 
\begin{align}
N_1 \le 2d \cdot L^{d-1} \cdot \bar{l}.
\end{align}
This last relation is because the vertices in $N_1$ must be near the boundary.  With $\bar{l}$ fixed, we note
\begin{align}
\lim_{L \to \infty} \frac{N_2}{N} = 1.
\end{align}
Then for $L$ large enough from (\ref{z-fbar-bound2}) one has
\begin{align}
\label{ineq}
\left( Z \left( \bar{f} \right) \right)^{\frac{1}{N}} \le \frac{1}{e} + \frac{\delta_1}{2}.
\end{align}
So for $\left| \Lambda \right|$ large enough we have the inequality we need (\ref{ineq}), with $\delta = \delta_3$.  The smaller $\left| \Lambda \right|$ (from this argument, and from that leading to $\Lambda_1$) are dealt with as in the discussion at the beginning of this step.

\end{document}